\documentclass[a4paper,11pt]{article}
\pdfoutput=1 

\usepackage{jcappub} 
\usepackage{subcaption}
\usepackage{graphicx}
\usepackage{caption}
\usepackage{afterpage}

\usepackage{bm}

\title{\boldmath Probabilistic learning for pulsar classification}

\author[a,b,1]{Sambatra Andrianomena,\note{Corresponding author.}}

\affiliation[a]{South African Radio Astronomy Observatory (SARAO), Black River Park, Observatory, Cape Town, 7925, South Africa}
\affiliation[b]{University of the Western Cape, Bellville, Cape Town 7535,
South Africa}

\emailAdd{andrianomena@gmail.com}

\abstract{In this work, we explore the possibility of using probabilistic learning to identify pulsar candidates. We make use of Deep Gaussian Process (DGP) and Deep Kernel Learning (DKL). Trained on a balanced training set in order to avoid the effect of class imbalance, the performance of the models, achieving relatively high probability of differentiating the positive class from the negative one (\textit{roc-auc} $\sim 0.98$), is very promising overall. We estimate the predictive entropy of each model predictions and find that DKL is more confident than DGP in its predictions and provides better uncertainty calibration. Upon investigating the effect of training with imbalanced dataset on the models, results show that each model performance decreases with an increasing number of the majority class in the training set. Interestingly, with a number of negative class $10\times$ that of positive class, the models still provide reasonably well calibrated uncertainty, i.e. an expected Uncertainty Calibration Error (UCE) less than $6\%$. We also show in this study how, in the case of relatively small amount of training dataset, a convolutional neural network based classifier trained via Bayesian Active Learning by Disagreement (BALD) performs. We find that, with an optimized number of training examples, the model -- being the most confident in its predictions -- generalizes relatively well and produces the best uncertainty calibration which corresponds to UCE = $3.118\%$.      
}

\begin{document}
\maketitle
\flushbottom

\section{Introduction}\label{sec:intro}

The detection of a pulsar PSR 1913+16 in a binary system by \cite{hulse1975discovery} and its subsequent monitoring \citep{taylor1982new} pointed toward the existence of gravitational radiation, an energy loss which is consistent with the decrease of the orbital period of the system. A couple of decades later, gravitational waves from a binary black hole merger were directly detected using the Laser Interferometer Gravitational-Wave Observatory (LIGO) experiment \citep{abbott2016observation}, confirming Einstein's prediction. PSR 1913+16, described as \textit{"an accurate clock moving at high velocity in the strong gravitational field of its unseen companion"} in \cite{taylor1982new}, was utilized as a laboratory test for gravity. Their results confirmed Einstein's General Relativity (GR) theory at 0.2\% level. Further tests using PSR B1534+12 \citep{stairs2002studies} and PSR J0737-3039A/B \citep{kramer2006tests} put constraints on GR at 0.7\% and 0.05\% level respectively. \cite{zhang2019constraints}, using ten binary systems \footnote{5 neutron star - neutron star and 5 neutron star - white dwarf systems.}, also investigated the viability of \textit{Screened modified gravity} by constraining the Post-Keplerian Parameters of the theory. These examples highlight the important role that pulsars play when addressing the validity of Einstein's GR and other alternative theories in highly non-linear regime.

In nuclear physics, \cite{pons2013highly} demonstrated the correlation between the inner crust composition of neutron star and its observed spin period, placing some constraints on the latter. \cite{kramer2016pulsars} listed some tests in fundamental physics where pulsars can be used as tools. \cite{siemens2019physics} showed that Pulsar Timing Arrays (PTAs) will help further our understanding of dark matter, are great probes for the detecting cosmic superstrings, and will be used to place constraints on gravitational wave spectrum in the inflationary universe. NANOGrav Collaboration \citep{arzoumanian2018nanograv, arzoumanian2020multi} is spending a great amount of effort in an attempt to detect stochastic gravitational-wave background using PTAs.

With the advent of upcoming big survey like SKA, a considerable increase of the number of  detected pulsars is expected. \cite{smits2009pulsar} estimated that SKA experiment would detect 14000 pulsars. At the time of writing, MeerKAT, a precursor of SKA-Mid, has observed 1005 pulsars \citep{bailes2020meerkat}. The huge amount of data from experiments requires an automated way to identify pulsar candidates. \cite{eatough2010selection}, for instance, trained an Artificial Neural Network (ANN) by considering 12 predictors as inputs\footnote{See their Table.~1.} to search for candidates and discovered a new pulsar using the method. Similar approaches, using 22 and 6 features as inputs to train an ANN was adopted in \cite{bates2012high} and \cite{morello2014spinn} respectively. Following \cite{morello2014spinn}, \cite{bethapudi2018separation} selected the same features (6 of them) to train neural network and tree based algorithms.
The Pulsar Image-based Classification System (PICS) prescribed by \cite{zhu2014searching} was more involved. Having highlighted the potential bias induced by the score based systems which use the predictor variables as inputs to the algorithms, \cite{zhu2014searching} opted for methods that extracted the features directly from the diagnostic plots. In other words, they fed the summed pulse profile (1D), time vs phase plot (2D image), frequency vs phase plot (2D image) and Dispersion Mesure (DM) curve (1D) to PICS, yielding an instance comprising 4 different inputs. The innovative approach consisted of two stages. In the first stage, each type of inputs was fed to two different methods, giving a total of 8 output scores which were then passed through a logistic regression for predictions in the second stage. They used convolutional neural network (CNN) combined with SVM to learn the features from the images and ANN (with dense layers) combined with SVM to extract features from 1D inputs.

Deep neural network classifiers tend to be ``\textit{overconfident}'' in their predictions which are point estimates by construction. It was shown in  \cite{nguyen2015deep, hein2019relu} that neural network classifiers are likely to classify an input, regardless of the fact that the latter is drawn from an out-of-distribution sample, with high probability (Softmax output). By estimating uncertainties, it is possible to assess how confident the classifier predictions are. Given a model with good uncertainty calibration, predictions associated with high predictive uncertainty can be discarded as they indicate what the model don't know. To estimate uncertainties in classification task, one can resort to probabilistic learning such as Bayesian Neural Network and Monte-Carlo (MC) Dropout model \cite{gal2016dropout}. 

In this work, we make use of Deep Gaussian Process (DGP) and Deep Kernel Learning (DKL) for pulsar classification. We also demonstrate the use of Bayesian Active Learning by Disagreement (BALD) to train a CNN classifier in the case where a relatively small amount of labeled data is available for training. Our aim is twofold: highlighting the predictive power of probabilistic models and estimating predictive uncertainty, which is done for the first time in pulsar classification, to the best of our knowledge. The paper is structured as follows: we present the dataset used in this work in Section~\ref{sec:dataset} and the methods we consider in Section~\ref{sec:algos}. The performance of the classifiers is presented in Section~\ref{sec:model-performance}. In Section~\ref{sec:uncertainty}, we provide the details of the predictive uncertainty estimation and assess the calibration of the uncertainties produced by the models. We investigate the effect of the imbalanced training dataset on how the models perform and the resulting prediction uncertainties in Section~\ref{sec:imbalance}. We show the use of BALD to classify pulsars in Section~\ref{sec:bald} and finally conclude in Section~\ref{sec:conclusion}.  

\section{Data}\label{sec:dataset}

\begin{figure}[tbp]
\centering
\includegraphics[width=0.8\textwidth]{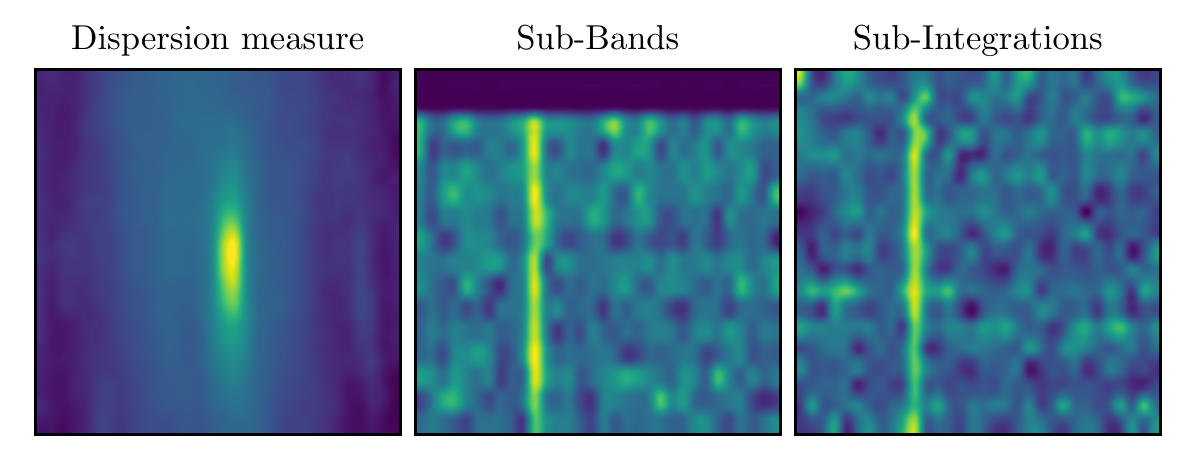}
\caption{Example of pulsar in the dataset we consider in this study. Left panel shows the 2D dispersion measure (DM), middle panel represents Sub-band and Sub-integration is shown on the right panel. Each input is composed of these three channels, mimicking RGB channels.}
\label{fig:pulsars}
\end{figure}

We consider pulsar data, named HTRU Medlat Data, which were collected by \cite{keith2010high} and \cite{morello2014spinn}.
The HTRU1 batched data, a subset of HTRU Data, contains 50000 labelled images for training and 10000 images for testing\footnote{Data can downloaded from \href{URL}{https://github.com/as595/HTRU1}.}. The imbalance ratio in both the training and testing dataset is about 1:49. Each input has three channels which are Dispersion Measure (channel 0), Sub-band (channel 1) and Sub-integration (channel 2). Each channel, which is a 2D image, has a resolution of $32\times32$ pixels. In our approach, we select all the channels -- Dispersion Measure, Sub-bands, Sub-integrations.
As an example, we present in Figure~\ref{fig:pulsars} the 2D map of each channel of a pulsar. Our task is a binary classification in which the positive class is pulsar and the negative class is non-pulsar (rfi). The target is categorical; $\bm{0}$ for non-pulsar and $\bm{1}$ for pulsar.

\section{Algorithms}\label{sec:algos}
In this section, we give an overview of the methods that are used in our analyses. We first present the approach used to find an optimal solution for each algorithm. In a probabilistic model inference, one can consider a sampling strategy such as Metropolis-Hastings algorithm 
\citep{metropolis1953equation, hastings70, chib1995understanding}
which is more accurate but time consuming as it fully explores the parameter space  or a faster method known as variational inference \citep{beal2003variational} but less accurate. We adopt the latter in our analyses.
\subsection{Stochastic variational inference (SVI)}
The approximation of the posterior probability emerges out of attempting to maximize the log marginal likelihood 
\begin{equation}\label{logmarg}
    {\rm log}\>p(\bf{y}) = {\rm log}\int {\rm d} \bm{\theta}\>\textit{p}({\bf y}, {\bm \theta}),  
\end{equation}
which, in general intractable, consists of averaging out all parameters in a model. Introducing a known variational distribution $q_{\gamma}$ in Eq.~\ref{logmarg} gives
\begin{equation}
     {\rm log}\>p(\mathbf{y}) = {\rm log}\int {\rm d}\bm{\theta}\>q_{\gamma}(\bm{\theta})\frac{p(\mathbf{y}, \bm{\theta})}{q_{\gamma}(\bm{\theta})}.
\end{equation}
By using Jensen's inequality which states that ${\rm log}\>\mathbb{E}[f(x)] \ge \mathbb{E}[{\rm log}\> f(x)]$ and introducing the posterior distribution yield
\begin{equation}
    {\rm log}\>p(\mathbf{y}) \ge \int {\rm d}\bm{\theta}\>q_{\gamma}(\bm{\theta}){\rm log}\>\frac{p(\bm{\theta}|\mathbf{y})p(\mathbf{y})}{q_{\gamma}(\bm{\theta})},
\end{equation}
where the right-hand side is known as the evidence lower bound ($\mathbb{ELBO}(q_{\gamma})$)\citep{wingate2013automated}. A little algebra gives
\begin{equation}
    {\rm log}\>p(\mathbf{y}) - \mathbb{ELBO}(q_{\gamma}(\bm{\theta})) = \int {\rm d}\bm{\theta}\>q_{\gamma}(\bm{\theta}){\rm log}\> \frac{q_{\gamma}(\bm{\theta})}{p(\bm{\theta} | \mathbf{y})},
\end{equation}
where the right-hand side is the Kullback-Leibler divergence ($\mathbb{KL}(q(\bm{\theta})||p(\bm{\theta}|\mathbf{y})$) , which measures the dissimilarity between two distributions. KL-divergence is both asymmetric\footnote{$\mathbb{KL}(q || p) \neq \mathbb{KL}(p || q)$} and positive, and as it gets closer to zero, the variational distribution ($q_{\gamma}(\bm{\theta})$) approaches the true posterior distribution. To approximate the latter, one chooses to either minimize the KL-divergence or maximize the evidence lower bound by varying the variational parameters $\gamma$ in a stochastic gradient descent (or ascent) manner. In all our analyses, we opt for maximizing the evidence lower bound.\\
\subsection{Deep Gaussian Process (DGP)}
As prescribed in \cite{damianou2013deep}, DGP consists of chaining up GP layers such that the outputs (latent spaces) of an intermediate layer are the inputs of the following one. Assuming we have two GP layers for simplicity, we have that

\begin{eqnarray}
\mathbf{y} & = & \bm{f}^{y}(\mathbf{t}) + \bm{\epsilon_{o}}, \\
\mathbf{t} & = & \bm{f}^{t}(\mathbf{X}) + \bm{\epsilon_{i}},
\end{eqnarray}
where $\bm{\epsilon_{i}}$ and $\bm{\epsilon_{o}}$ are Gaussian noise at the hidden layer and output layer respectively, $\mathbf{t}$ is a noisy realization of the intermediate latent function $\bm{f}^{t}(\mathbf{X}) \sim \mathcal{GP}(\mathbf{0}, K^{t}(\mathbf{X}, \mathbf{X'}))$ and the observation $\mathbf{y}$ is also a noisy realization of the latent function $\bm{f}^{y}(\mathbf{t}) \sim \mathcal{GP}(\mathbf{0}, K^{y}(\mathbf{t}, \mathbf{t'}))$. We refer the interested reader to \cite{damianou2013deep} for full details of the theory. In practice, we consider a DGP composed of two layers of variational sparse GPs (see \ref{gaussian_process} for more details), each with a radial basis function (RBF) kernel. For the training we select: Adam optimizer, learning rate = 0.02, batch size = 256 and number of inducing points = 64. The algorithm is trained by maximizing the $\mathbb{ELBO}$ via variational inference for 31 epochs. It is noted that due to the stochastic nature of the prediction, owing to the hidden layer, we sample 100 predictions in each forward pass. Both the predictive mean $\hat{\bm{f}}$ and standard deviation $\hat{\bm{\sigma}}$ of the latent function (logit) are defined as
\begin{equation}
    \hat{\bm{f}} = \frac{1}{N} \sum_{i = 1}^{N} \bm{f}_{i},  
\end{equation}
\begin{equation}
    \hat{\bm{\sigma}} = \frac{1}{N} \sum_{i = 1}^{N} \bm{\sigma}_{i},  
\end{equation}
where $\bm{f}_{i}$ and $\bm{\sigma}_{i}$ are the mean and standard deviation in each sample respectively and $N$ is the sample size in each forward pass.
The implementation of this method is achieved with {\sc Pyro} library 
\cite{bingham2019pyro}. 

\subsection{Deep Kernel Learning (DKL)}

For inputs with high dimensional features, e.g. 2D image with $256\times 256$ pixels, using kernel based classifier might be a challenge. \cite{wilson2016stochastic} exploited the capacity of a deep neural network and the flexibility of a GP to arrive at a probabilistic deep network which they named Stochastic Variational Deep Kernel Learning. The salient features from the image (input) are first extracted via a neural network model. 
\begin{table}[h!]
 \centering
 \begin{tabular}{|c|c|c|}
  \hline
   & Layer & (in channel, out channel, kernel, stride) \\
  \hline
  1  & Convolutional Layer & (3, 16, 3$\times$3 , 2)\\[2pt]
  2  & ReLU Activation & --\\[2pt]
  3  & Convolutional Layer & (16, 16, 3$\times$3 , 1)\\[2pt]
  4  & ReLU Activation & --\\[2pt]
  5  & Convolutional Layer & (16, 32, 3$\times$3 , 2)\\[2pt]
  6  & ReLU Activation & --\\[2pt]
  7  & Convolutional Layer & (32, 32, 3$\times$3 , 1)\\[2pt]
  8  & ReLU Activation & --\\[2pt]
  9  & Convolutional Layer & (32, 32, 3$\times$3 , 2)\\[2pt]
  10  & ReLU Activation & --\\[2pt]
  11  & Convolutional Layer & (32, 32, 3$\times$3 , 1)\\[2pt]
  12  & ReLU Activation & --\\[2pt]
  13 & Flatten & --\\[2pt]
  14 & $\rm{Fully\ Connected\ Layer}$ &  (512, 1024, --, --)\\[2pt]
  15 & ReLU Activation & --\\[2pt]
  \hline
 \end{tabular}
 \caption{The architecture of the feature extractor that is considered in this study.}
 \label{table:feature_extractor}
\end{table}
The extracted features are fed into a Gaussian process which in turn outputs the predictive mean $\hat{\bm{f}}$ and standard deviation $\hat{\bm{\sigma}}$ of the latent function. Predicting a class label\footnote{In the case of classification.} is done by feeding a sample ${\bm{f}} \sim \mathcal{N}(\hat{\bm{f}}, \hat{\bm{\sigma}})$ to a sigmoid function. The weights of the network together with the GP hyperparameters are learnt by maximizing $\mathbb{ELBO}$ using variational inference. In this study, we consider the architecture presented in Table~\ref{table:feature_extractor} as our feature extractor and the base kernel of the GP layer is also RBF. For the training we choose: Adam optimizer, learning rate = 0.001, batch size = 256 and number of inducing points = 64. The training converges over 200 epochs. We also use {\sc Pyro} for the model implementation.

\section{Model performance}\label{sec:model-performance}
As a way to mitigate the effect of imbalance on the training, we consider a representative sample which is composed of all positive instances in the original training dataset\footnote{Which has 50000 examples in total.} and the same number of randomly drawn negative instances, giving a balanced dataset of 1990 instances in total. That latter is split into training set (80$\%$) and validation set (20$\%$). The original testing dataset\footnote{Which has 10000 examples.} has 199 positive instances (pulsar) which are combined with randomly drawn negative instances (rfi) from the same testing dataset to get a balanced test set with 398 examples in total. It is worth noting that, although we investigate the effect of imbalanced training dataset on the predictions in the following section, we defer a thorough investigation on dealing with imbalance classification for future work.
\begin{table}[h!]
 \centering
 \begin{tabular}{|c | c | c | c | c | c |}
  \hline
   & $f_{1}$\textit{-score} & \textit{recall} &\textit{precision} & \textit{roc-auc} & \textit{specificity}\\
  \hline
  DGP & 0.955 & 0.924 & 0.989 & 0.992 & 0.989\\[2pt]
  \hline
  DKL & 0.969 & 0.954 & 0.984 & 0.992 & 0.984\\[2pt]
  \hline
  DCGAN-L2-SVM-2\cite{guo2019pulsar}& 0.964 & 0.963 & 0.965 & -- & -- \\[2pt]
  \hline
  DCNN-S \cite{wang2019pulsar} & 0.962 & 0.962 & 0.963 & -- & -- \\[2pt]
  \hline
  BALD & 0.947 & 0.914 & 0.983 & 0.987 & 0.984 \\[2pt]
  \hline
 \end{tabular}
 \caption{Performance metrics used to assess the performance of the classifiers.}
\label{table:metric_results}
\end{table}
To assess the performance of each method, we use various metrics
\begin{itemize}
    \item \textit{recall} (also known as \textit{sensitivity} or \textit{completeness}) encodes the minimization of the number of false negatives which are positive instances misidentified as negative ones.
    \item \textit{specificity} indicates how well the number of false positives, which are negative instances incorrectly classified as positive instances, is minimized.
    \item \textit{precision} (also known as \textit{purity}) denotes how well the positive class is identified.
    \item $f_{1}$-\textit{score} is simply the harmonic mean of the \textit{recall} and \textit{precision}. 
    \item \textit{roc-auc} is the degree of separability which indicates how good a classifier performs in terms of making the distinction between the two classes in our case which is a binary classification.
\end{itemize}
We refer the interested reader to \cite{andrianomena2020classifying} for a more explicit summary of the metrics mentioned above. Table.~\ref{table:metric_results} shows the results corresponding to each algorithm. Overall, the values of the metrics are  $> 0.92$ indicating a relatively good performance of all the methods. DKL, with its higher \textit{recall}, is more sensitive than DGP. However comparing the values of \textit{recall} with those of \textit{specificity} suggests that all learners are more likely to misclassify pulsars. This can be explained by the fact that the salient features of pulsar candidates can be challenging to extract due to the strength of the signal for instance. Results also show that the models exhibit similar performance in terms of identifying the positive class, as evidenced by the values of their \textit{precision}.
In pulsar candidate search, provided the relative scarcity of the object, the idealized scenario is a classifier that has high \textit{sensitivity} and a great capability of detecting the positive class (high \textit{precision}). However, the \textit{precision/recall} tradeoff which is known as the effect of optimizing one to the detriment of the other, may be challenging to overcome. Therefore, even though optimizing \textit{recall} is the main objective, it can only be done at a fixed value of tolerable \textit{precision}\footnote{And vice versa.}. A very useful metric that encodes the effect of that tradeoff is $f_{1}$-\textit{score} which is in favour of a method that optimizes both \textit{precision} and \textit{recall}. 

Apart from \textit{roc-auc} score, the other metrics considered in this work are all based on a single value of a threshold which can be a score (e.g. logit) or a probability. The predicted class label is positive or negative whether the predicted probability/score is above or below the threshold respectively. In our case, the value of that threshold, which is a probability, is 0.5\footnote{It is the default value in most cases.}. Therefore, depending on the problem, the value of the threshold can be adjusted to meet a target value of a metric particularly chosen for the task. In general, the \textit{roc-auc} score, which is the area under the curve of the \textit{true positive rate}\footnote{Another name for \textit{recall}.} against the \textit{false positive rate}\footnote{This is defined as \textit{false positive}/(\textit{false positive} + \textit{true positive})}, is commonly used to compare how well different methods generalize in a given classification task. The reason is that the curve highlights the tradeoff between \textit{recall} and \textit{false positive rate}, in other words the value of the former that corresponds to that of the latter at all possible values of the threshold. Therefore \textit{roc-auc} can be viewed as the mean value of the \textit{recall} at different values of the threshold. High values of \textit{roc-auc} ($\sim 0.99$) indicate that the capability of the models to distinguish pulsars from non-pulsars is promising.

In Table.~\ref{table:metric_results}, we show some results from previous studies. \cite{guo2019pulsar} made use of the features extracted from a generative model (Deep Convolutional Generative Adversarial Network) trained on the HTRU dataset as inputs to a support vector machine (SVM). They fed the Sub-bands and Sub-Integrations maps into their model for the learning process and achieved good performance as suggested by their chosen metric values all above 0.96. \cite{wang2019pulsar}, who also selected Sub-bands and Sub-Integrations maps as inputs, resorted to data augmentation to train their CNN and obtained a performance similar to the model used in \cite{guo2019pulsar}. Overall our methods, although slightly less sensitive (especially DGP), perform comparably to those of \cite{guo2019pulsar} and \cite{wang2019pulsar}. However, since we consider different inputs for our models and use slightly different dataset, the idea behind this comparison is to check that our results are consistent with those of previous studies.

\section{Quantifying uncertainty}\label{sec:uncertainty}
Uncertainty in deep learning was shown to have two main components \cite{gal2016uncertainty, kendall2017uncertainties}. Epistemic uncertainty, also referred to as model uncertainty, describes the uncertainty in the model parameters and is propagated through to the model predictions. The model ignorance is encoded in this type of uncertainty and decreases with more data for the training. The other component is aleatoric uncertainty which encodes the inherent noise in the observed data. The aleatoric uncertainty which cannot be mitigated with more data, can be dependent (heteroscedastic) or independent (homoscedastic) of the inputs. In our analyses, we estimate the predictive entropy (total uncertainty) which is the sum of the epistemic and aleatoric uncertainties.
\subsection{Predictive uncertainty}
We make use of the prescription described in \cite{gal2016uncertainty, kendall2017uncertainties} and also followed by \cite{mohan2022quantifying} in their in-depth investigation of deep learning uncertainty in radio galaxy classification. At a test point, the predictive entropy is given by 
\begin{equation}\label{eq:predictive-entropy}
    \mathcal{H}(y_{*} | X_{*}, \mathcal{D}) = -\sum_{c = 0}^{C-1} p(y_{*} = c | X_{*}, \mathcal{D})\>{\rm log}\>p(y_{*} = c | X_{*}, \mathcal{D}), 
\end{equation}

where $C$ is the number of classes and the predictive probability is given by

\begin{equation}
    p(y_{*} = c | X_{*}, \mathcal{D}) = \frac{1}{N}\sum_{i = 1}^{N}{\rm Softmax}(\bm{f}_{i}(X_{*}))
\end{equation}

where $\bm{f}_{i}(X_{*})$ is a sample that can be obtained from a stochastic forward pass in the case of MC Dropout model \cite{gal2016dropout}. In the case of Bayesian Neural Network, $\bm{f}_{i}(X_{*})$ is a prediction obtained by sampling from the posterior distribution of the model weights. In our case\footnote{Both DGP and DKL.}, provided that our models output posterior mean and uncertainty of the logit ($\hat{\bm{f}}$, $\hat{\bm{\sigma}}$), $\bm{f}_{i}$ can be sampled from the distribution $ \mathcal{N}(\hat{\bm{f}}, \hat{\bm{\sigma}})$. 
\begin{figure}[h!]
\centering
\includegraphics[width=.6\textwidth]{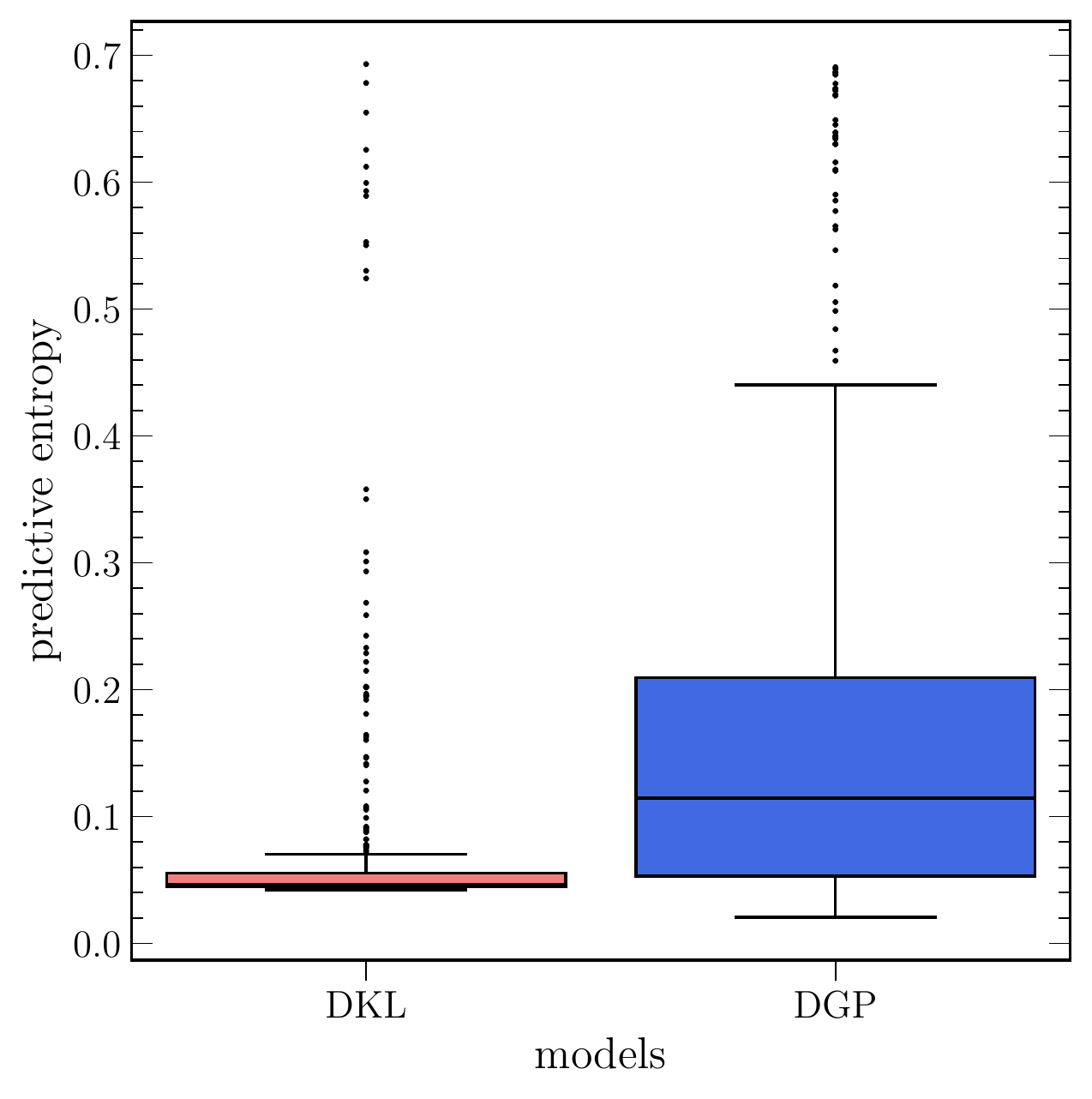}
\caption{Distribution of the predictive uncertainty. Red denotes the uncertainty distribution corresponding to DKL model, whereas blue corresponds to that of DGP model.}
\label{fig:uncertainty_quantification}
\end{figure}
In information theory, the unit of entropy depends on the base of logarithm which is used in Equation~\ref{eq:predictive-entropy}. We use natural logarithm throughout, therefore the unit is natural unit (nat). The entropy is maximum (corresponding to maximum uncertainty) if predicting rfi and predicting pulsar are equally likely\footnote{In that case, $p(y_{*} = 0|X_{*}) = p(y_{*} = 1|X_{*}) = 0.5$, hence $\mathcal{H}$ = $2\times{\rm log(0.5)}\times 0.5 = 0.693147$.}. 

We show the distributions of predictive uncertainty obtained from each model in Figure~\ref{fig:uncertainty_quantification}. The red and blue boxes correspond to the uncertainty distribution of DKL and DGP respectively. Each box indicates the first quartile ($Q_{1}$), the median, the third quartile ($Q_{3}$). The minimum and maximum \footnote{Which are denoted by the whiskers.} are given as a function of the interquartile range ($IQR$), and the dots denote the outliers.
The results suggest that DKL model has more confidence in its predictions compared to DGP. This is evidenced by both its lower median value of entropy (DKL: 0.046 nats, DGP: 0.114 nats) and its smaller $IQR$.
This can be explained by the capacity of DKL model which uses convolutional layers to capture the relevant features of the input, unlike the DGP model whose kernel is built from high dimensional data\footnote{Each input has to be flattened to get a one dimensional array with length $3\times32\times32$.}. 

\subsection{Uncertainty calibration}
\begin{figure}[h!]
\centering
\includegraphics[width=.6\textwidth]{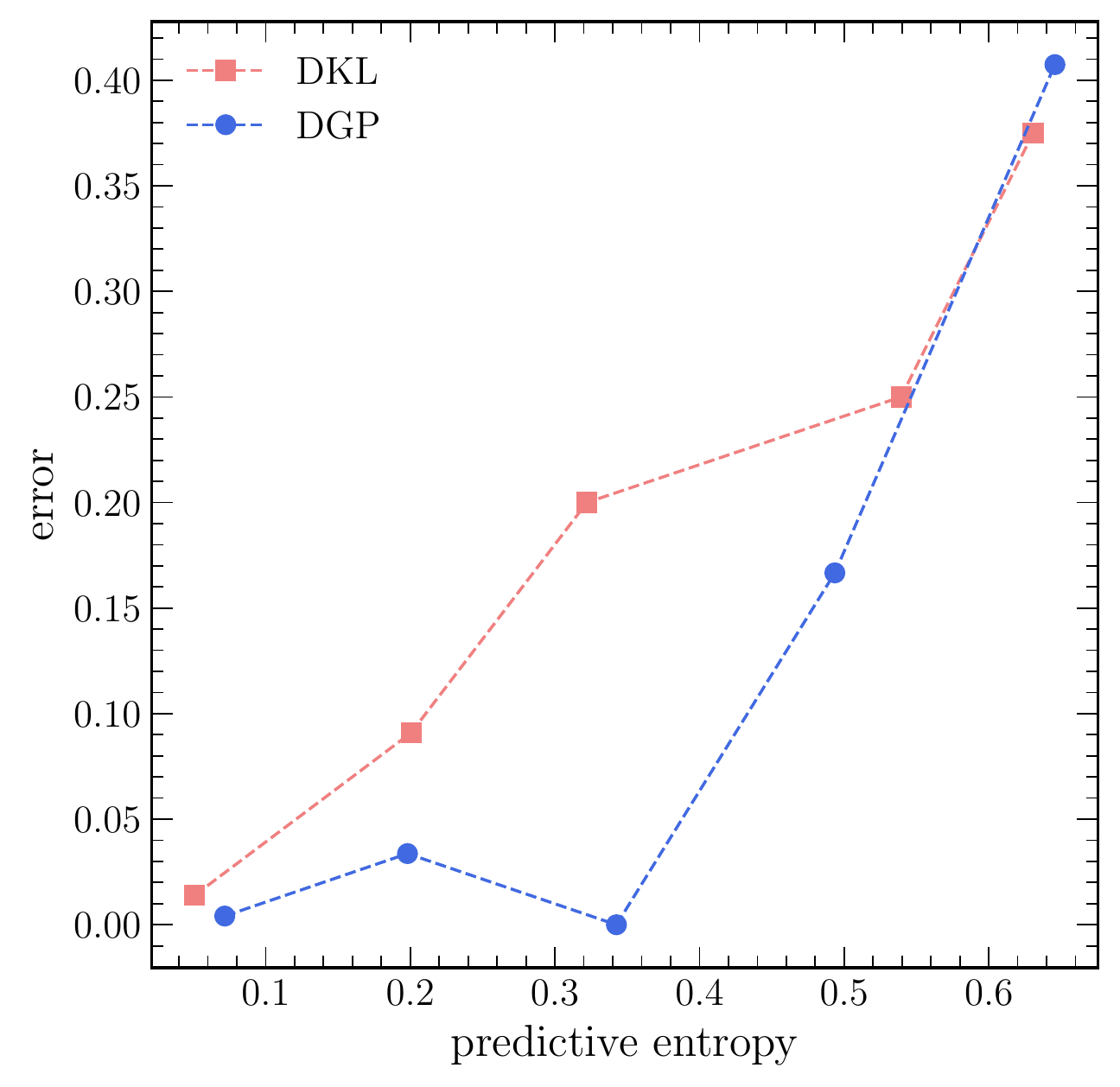}
\caption{Test error in each bin as a function of average of predictive uncertainties in the same bin. }
\label{fig:uncertainty_calibration}
\end{figure}
The uncertainty estimated from our models which are trained via variational inference can be miscalibrated. As defined in \cite{laves2019well}, the uncertainty is perfectly calibrated if
\begin{equation}\label{eq:calibration}
    \mathbb{P}(\hat{y}\ne y | \mathcal{H}(y | X, \mathcal{D}) = q) = q,\> \forall q \in [0,1].
\end{equation}

In other words, the test error\footnote{Or top-1 error in the case of multi-class.}in a test set ($X, y$) is strongly correlated with the predictive uncertainty in the case of perfect calibration. To assess the calibration of the uncertainty obtained from our model, we compute the expected Uncertainty Calibration Error (UCE), which was prescribed by \cite{laves2019well}, and is given by
\begin{equation}\label{eq:uce}
    {\rm UCE} = \sum_{m = 1}^{M}\frac{|B_{m}|}{n}|{\rm err}(B_{m}) - {\rm uncert}(B_{m})|,
\end{equation}
where $M$ is the number of bins, $B_{m}$ denotes all the instances in a given bin, $n$ is the total number of instances, ${\rm err}(B_{m})$ indicates the average error in each bin which is defined as
\begin{equation}\label{eq:average-error}
    {\rm err}(B_{m}) = \frac{1}{|B_{m}|}\sum_{i \in B_{m}} {\bf 1}(\hat{y}_{i} \neq y),
\end{equation}
and the average uncertainty in each bin ${\rm uncert}(B_{m})$ is given by
\begin{equation}
    {\rm uncert}(B_{m}) = \frac{1}{|B_{m}|}\sum_{i \in B_{m}} {\rm uncert}_{i}.
\end{equation}
${\rm uncert}_{i}$ is the predictive entropy in our case. Due to the relative small number of examples in the test set, we set the number of bins to 7.
We present in Figure~\ref{fig:uncertainty_calibration} the variation of test error as a function of the average of predictive entropy. Red squares and blue circles are the errors (see Equation~\ref{eq:average-error}) obtained respectively from DKL and DGP in each bin. It is noticed that there are only 5 data points for each model. This is due to the fact that some bins are empty. Using Equation~\ref{eq:uce}, we obtain ${\rm UCE} = 4.897\%$ and ${\rm UCE} = 12.728\%$ with DKL and DGP respectively. This suggests that DKL produces good uncertainty calibration and the uncertainty from DGP model is not well calibrated. It is also possible to assess the strength of the correlation between the error and the predictive entropy (shown in Figure~\ref{fig:uncertainty_calibration}) by computing the Pearson's correlation coefficient but due to the relatively small amount of data, that quantity is highly dependent on the number of bins selected and hence will not be considered in our analyses.  

Another approach, which was considered by \cite{li2021deep} to check whether a model can produce reliable uncertainty, is to compare the variance of the ratio of true positive/true negative ($\sigma_{TP/TN}^{2}$) with that of the ratio of false positive/false negative ($\sigma_{FP/FN}^{2}$). A model produces good uncertainty if $\sigma_{FP/FN}^{2} \gg \sigma_{TP/TN}^{2}$. As presented in Table~\ref{tab:uncertainty_ratios}, $\sigma_{FP/FN}^{2}$ is about two order of magnitude greater than $\sigma_{TP/TN}^{2}$ for all models, suggesting that the classifiers give good uncertainty estimation. 
\begin{table}[h!]
\centering
\begin{tabular}{|c | c | c |}
\hline
  & $\sigma_{TP/TN}^{2}$ & $\sigma_{FP/FN}^{2}$ \\
 \hline
 DGP & $9.110\times10^{-5}$ & $7.189\times10^{-3}$\\[2pt]
 \hline
 DKL & $5.107\times10^{-5}$ & $4.626\times10^{-3}$\\[2pt]
 \hline
\end{tabular}
\caption{Variance of the ratio of true positive/true negative, and that of false positive/false negative for each model.}
\label{tab:uncertainty_ratios}
\end{table}
The results in Table~\ref{tab:uncertainty_ratios} also indicate that uncertainty estimation from DKL model is more reliable.
%

\section{Imbalance classification}\label{sec:imbalance}
In this section, although dealing with imbalance is outside the scope of this study, we address the effect of the imbalanced training dataset on both the performance of the classifiers and the uncertainty estimation. 
To this end, we consider three different imbalance ratios of the training set $\rho$ = 1:1, 1:10, 1:49\footnote{This is the imbalance ratio of the original training set.}. 
\subsection{Effect on prediction}
\begin{figure}[h!]
\centering
\includegraphics[width=1\textwidth]{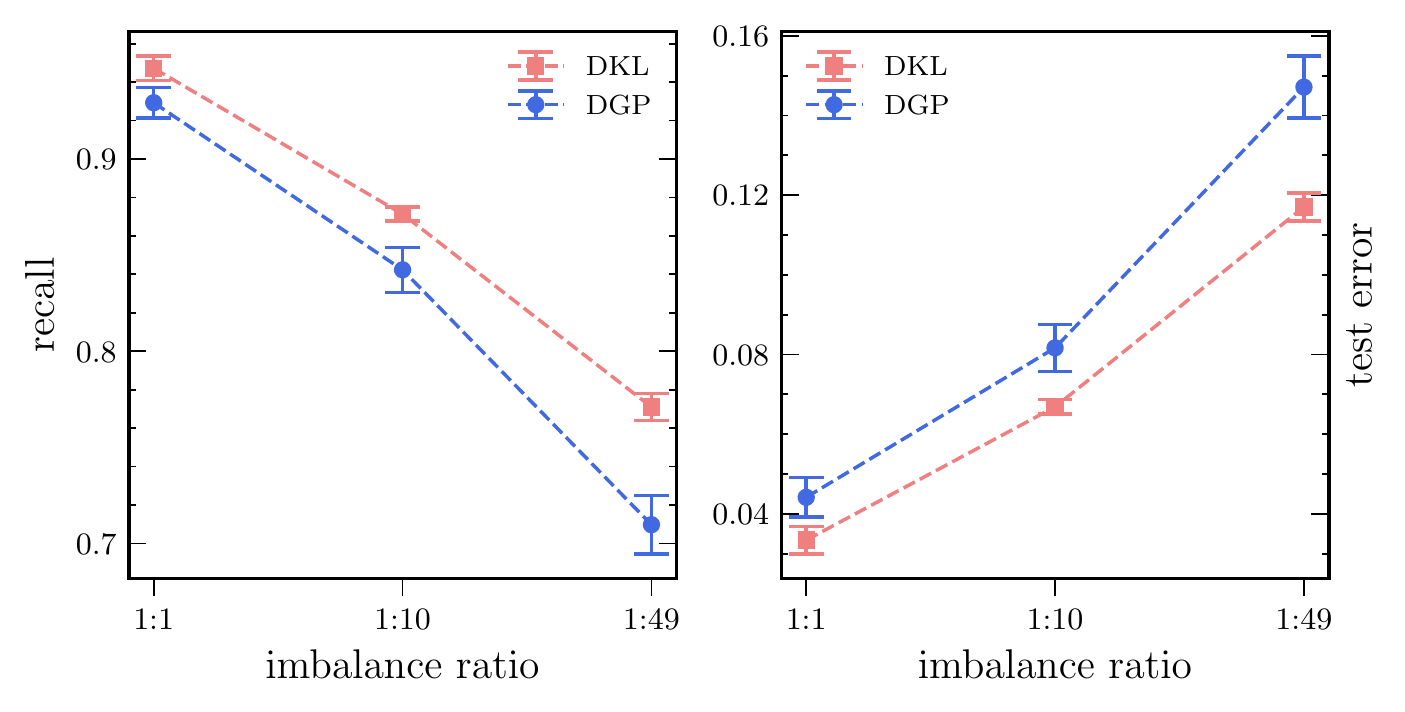}
\caption{Effect of training with imbalanced dataset on \textit{sensitivity}, test error.}
\label{fig:imbalance_recall_testerror}
\end{figure}

In order to check the effect of the imbalance on the performance of the classifiers, they are tested on the same balanced test set as in Section~\ref{sec:model-performance} after being trained on three different datasets, each with a specific $\rho$. To obtain the results in Figure~\ref{fig:imbalance_recall_testerror}, we sample 200 predictions for each instance of the balanced test set (396 instances) in each scenario\footnote{Training with a dataset with a specific imbalance ratio.}. 
We first investigate the influence of $\rho$ on the \textit{recall} of the learners. The \textit{sensitivity} of the models decreases with an increasing imbalance ratio as shown in the left panel in Figure~\ref{fig:imbalance_recall_testerror}. This can be accounted for by the fact that the larger number of negative classes in the training dataset causes the models to be biased against the positive ones at test time. The bias gets stronger with larger number of the majority class. For DGP model, the standard deviation of the predictions increases with $\rho$. Whereas the standard deviation of the DKL predictions is less sensitive to the increasing number of the majority class in the training sample. Similar trend is also observed with the test error as a function of the imbalance ratio (see right panel in Figure~\ref{fig:imbalance_recall_testerror}). Both models are more prone to error in their predictions with higher imbalance ratio. The standard deviation of the test error from DGP predictions is more impacted by the bias of the training set.

\subsection{Effect on uncertainty}
\begin{figure}[h!]
\centering
\includegraphics[width=1\textwidth]{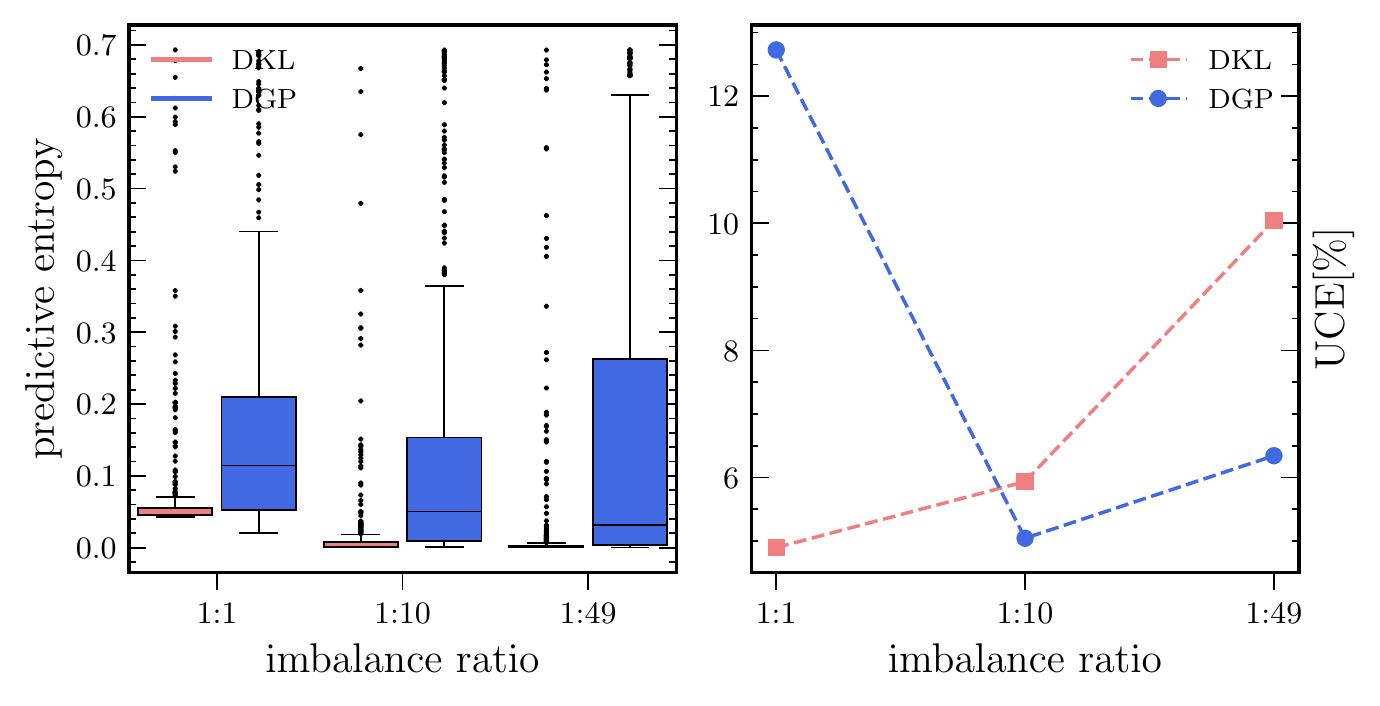}
\caption{Effect of training with imbalanced dataset on the predictive entropy. \textit{Left}: distribution of the predictive entropy related to each model for each imbalance ratio. \textit{Right}: uncertainty calibration UCE corresponding to each model as a function of the imbalance ratio.}
\label{fig:imbalance_predictive_entropy}
\end{figure}

In a similar way, we also analyze the effect of the bias on the predictive uncertainty by considering the three cases with different imbalance ratio. The left panel of Figure~\ref{fig:imbalance_predictive_entropy} shows the variation of the distribution of the uncertainty as a function of $\rho$, and the resulting calibration uncertainty for each case is presented on the right panel of Figure~\ref{fig:imbalance_predictive_entropy}.
Using the same balanced test set, it appears that the uncertainty produced by DKL model improves as the number of majority class increases in the training data. This is indicated by smaller \textit{IQR} and lower median value as $\rho$ increases. However, the uncertainty calibration obtained from DKL model degrades as a function of an increasing imbalance ratio of the training dataset. This implies that although the uncertainty seemingly gets better with the increasing number of majority class in the training data, its calibration gets worse. When considering the whole dataset (50000 instances, with $\rho$ = 1:49) to train DKL model, the resulting UCE on the balanced test set reaches $10\%$. 

For the DGP model, the median value of the distribution of its corresponding predictive entropy decreases as the number of the negative classes in the training set increases but the \textit{IQR} gets smaller when the imbalance ratio of the training data is 1:10 and is the largest when the model is trained on the whole dataset (1:49) (see left panel in Figure~\ref{fig:imbalance_predictive_entropy}). Interestingly, the uncertainty produced by DGP is well calibrated (UCE = $5.044\%$) when the bias $\rho$ = 1:10, but the calibration relatively degrades (UCE = $6.342\%$) when the DGP model is trained on the entire dataset.   
\section{BALD}\label{sec:bald}
In general, in order to achieve good generalization in the case of image classification, neural network needs to be trained with a substantial amount of the data, e.g. MNIST dataset \cite{lecun1998mnist} has 60000 instances. In production, labeling examples that will be used to train neural networks for a given task, may be expensive and active learning can be used to train the models with an optimized number of instances while achieving good generalization. In this section, we demonstrate the use of active learning approach to classify pulsar candidates, given the relatively small size of the training set.  

Assuming there is a relatively large dataset with only a small number of labeled data points. In active learning, the training starts with that small amount of data. At the end of one step, which amounts to training a model for a specific number of epochs, an acquisition function selects a number of unlabeled data points from the pool\footnote{The large set of unlabeled data.} and asks an ``$\textit{oracle}$" to label them. The newly labeled data points are added to the training set used in the previous active learning step and a new step starts with the updated training set. The amount of training set increases at each step until convergence. There are many active learning approaches but we make use of Bayesian active learning \cite{gal2017deep, atighehchian2020bayesian} which is scalable to both high dimensional data and large dataset. 

In general, the ``$\textit{oracle}$" that is requested to label the queried data points from the pool is a human expert, but in our work the data points are already labeled such that they are retrieved from the pool using the indices which are selected by the acquisition function. To query new data points from the pool, we consider Bayesian Active Learning by Disagreement (BALD) \cite{houlsby2011bayesian} as acquisition function. The new points that are selected are those for which the mutual information between predictions and model posterior are maximized, according to
\begin{equation}\label{eq:bald}
   X^{*} = {\rm argmax}_{X \in \mathcal{D}_{pool}} \mathcal{I}[y, \bm{\theta}|X, \mathcal{D}_{train}], 
\end{equation} 
where $\mathcal{D}_{pool}$ and $\mathcal{D}_{train}$ are the sets of points in the pool and a training set at a given step respectively. The mutual information is given by \cite{houlsby2011bayesian} 
\begin{equation}\label{eq:mutual_information}
    \mathcal{I}[y, \bm{\theta}|X, \mathcal{D}_{train}] = \mathcal{H}[y|X, \mathcal{D}_{train}] - \mathbb{E}_{\bm{\theta} \sim p(\bm{\theta}|\mathcal{D}_{train})}[\mathcal{H}[y|X,\bm{\theta}]].
\end{equation}
The mutual information in Equation~\ref{eq:mutual_information} can also be used to estimate the epistemic uncertainty \cite{gal2016uncertainty}, such that the selection criterion in Equation~\ref{eq:bald} can be interpreted as a way to seek for $X^{*}$ that maximize the epistemic uncertainty. 

In our setup,  we split the balanced data (1990 examples) into $80\%$ and $20\%$ which constitute the pool and validation set respectively. The initial dataset for the first active learning step is composed of 100 instances from the the pool and at each subsequent step 30 points with the maximum mutual information are queried by the acquisition function from the pool and added to the training set for the next step. At each step, the model is trained for 20 epochs and we consider a batch size of 16. The number of active learning steps is 5, such that the total number of data points used for the training is 220\footnote{This is given by 100 + $4\times30$.} which amounts to only about $14\%$ of the pool data. The model considered for the active learning process is similar to the feature extractor used in DKL model, but batch normalization is added after each convolutional layer before ReLU activation and another dense layer is added to output the predictions. Another key component is the dropout layer with a probability of 0.2 before the output layer as it plays a crucial role in computing the uncertainty during the training (MC Dropout model \cite{gal2016dropout}). For the implementation, we use {\sc{Baal}} library \cite{atighehchian2020bayesian}. 
\begin{figure}[h!]
\centering
\includegraphics[width=0.6\textwidth]{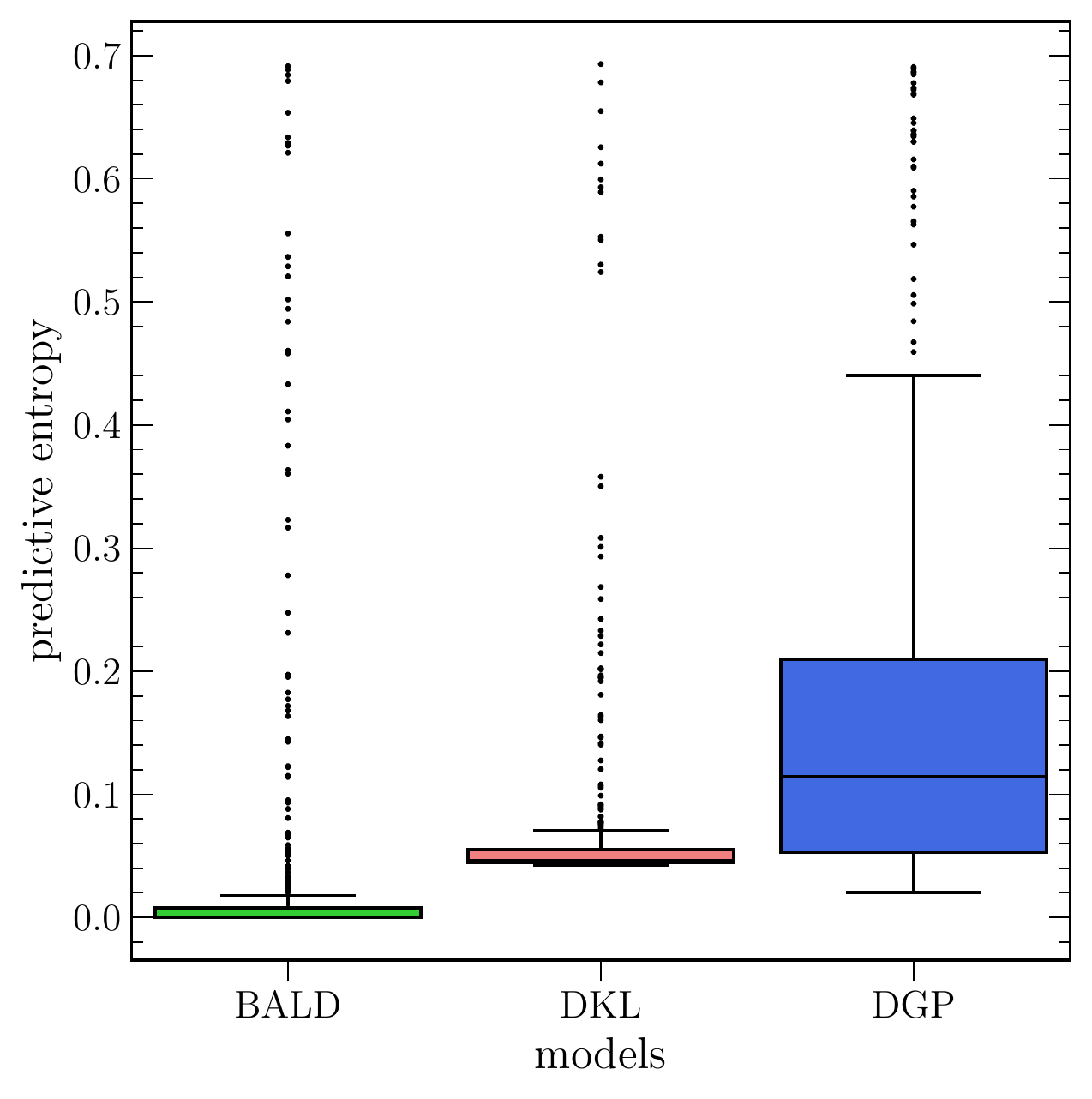}
\caption{Comparing the predictive entropy obtained from all models.}
\label{fig:all_models_predictive_entropy}
\end{figure}

The results obtained from using the BALD method to train the neural network based classifier is shown in Table~\ref{table:metric_results}. Overall the performance of the BALD method is comparable with those of DGP and DKL models, although its \textit{sensitivity} is relatively lower. However, provided the amount of data used to train the model, the results look very promising. By construction, the uncertainty can be estimated using the dropout layer during test time. To obtained the predictive entropy distribution, shown in Figure~\ref{fig:all_models_predictive_entropy}, we also run 200 forward passes. As indicated in Figure~\ref{fig:all_models_predictive_entropy}, the model trained with BALD is more confident about its predictions, i.e. lower median value and smaller \textit{IQR} compared to DGP and DKL, and its corresponding UCE, which is the lowest (see Table~\ref{tab:uncertainty_calibration_allmodels}), indicates that the model produces well calibrated uncertainty.    

\begin{table}[h!]
\centering
\begin{tabular}{|c | c |}
\hline
  & UCE $\left[\%\right]$  \\
 \hline
 DGP & 12.728 \\[2pt]
 \hline
 DKL & 4.897 \\[2pt]
 \hline
 BALD & 3.118 \\[2pt]
 \hline
\end{tabular}
\caption{Uncertainty Calibration Error obtained from all models.}
\label{tab:uncertainty_calibration_allmodels}
\end{table}
\section{Conclusion}\label{sec:conclusion}
This work has demonstrated the use of deep probabilistic learning to identify pulsars based on its features composed of 2D dispersion measure, sub-bands and sub-integrations. To this end, we have made use of a Deep Gaussian Process (DGP) which comprises two layers of Gaussian Process, a Deep Kernel Learning (DKL) which utilizes  
a neural network capacity to extract the features to be fed to the kernel of a Gaussian Process layer. Apart from highlighting the predictive power of the probabilistic methods used in this study, we have also estimated the uncertainty in their predictions and showed how good the calibration of the uncertainty is. The original dataset is strongly biased against the positive class with an imbalance ratio of $\rho$ = 1:49. To deal with the potential impact of the bias on the models, we have performed undersampling on the majority class (via random sampling) and the models have been trained using well balanced dataset. Nevertheless, we have analyzed the effect of the imbalance ratio on the performance of the classifiers and the resulting predictive uncertainty. Provided that the number of pulsars available for training is relatively small in general, which can pose an issue for deep network to generalize well, we have explored the possibility of using a convolutional neural network based classifier which is trained via active learning whose acquisition function is Bayesian Active Learning by Disagreement (BALD). 

The two models, DGP and DKL, exhibit great capability of differentiating pulsars from non-pulsars as indicated by their \textit{roc-auc} $> 0.98$. The small difference in their \textit{recall} suggests that DGP is more likely to misclassify the positive class but overall the classifiers generalize well. The performance of our methods is comparable to that achieved in previous studies. The lower median value of its predictive uncertainty distribution (0.046 nats) and the smaller interquartile range (\textit{IQR}) indicate that DKL shows more confidence in its predictions compared to DGP model which has larger \textit{IQR} and higher median value of predictive entropy distribution (0.114 nats). This can be attributed to DKL's higher capacity, which is achieved by utilizing convolutional neural network (CNN) based feature extractor to encode the salient features from the inputs. To assess the calibration of the uncertainty produced by the models, the expected Uncertainty Calibration Error (UCE) has been used. Results show that the uncertainty produced by DKL model, achieving UCE = $4.897\%$, is better calibrated than that estimated by DGP with UCE = $12.738\%$. The comparison between the variance of the ratio of true positive/true negative $\sigma^{2}_{FP/FN}$ with that of the ratio of false positive/false negative $\sigma^{2}_{FP/FN}$ is also considered to evaluate the quality of the uncertainty. For all models, $\sigma^{2}_{FP/FN}$ is larger than $\sigma^{2}_{TP/TN}$ by about two orders of magnitude overall, indicating that 
reliable uncertainty can be obtained from the models.

By considering three scenarios, each with a given bias $\rho$ = 1:1, 1:10, 1:49 of the training dataset, we have found that the predictive power of the models (DKL and DGP) degrades with an increasing imbalance ratio of the training dataset, as shown by a decreasing \textit{recall} and an increasing test error with an increase in the imbalance ratio. It has been found that as the number of majority class in the training dataset increases, DKL model becomes more confident in its predictions (lower median value of predictive entropy distribution and smaller \textit{IQR}), 
whereas the resulting uncertainty calibration gets poorer (increasing UCE). Interestingly, DGP model appears to produce well calibrated uncertainty with a relatively small imbalance $\rho$ = 1:10 in the training dataset. However with $\rho$ = 1:49, the \textit{IQR} of the resulting predictive uncertainty 
distribution is the largest, and the corresponding UCE reaches $6.324\%$. It can be argued that both DKL and DGP can still perform reasonably well when trained on dataset in which the number of majority class is less than $10\times$ that of the minority class. 

Using BALD method, a CNN based classifier with a dropout layer is trained with only using 220 instances in order to achieve generalization. This optimized number of training data points results from the learner requesting, via the acquisition function, data points with maximum mutual information (or epistemic uncertainty) to be included in the training dataset for the next active learning steps. As a whole, the performance of the CNN model obtained via the BALD approach is comparable to that of the other two models (DKL and DGP). Its \textit{recall}, which is relatively lower (0.914), denotes that it is more likely to misclassify the positive class. Compared to the other models, the CNN model trained via BALD is the most confident in its predictions, as shown by the median value (the lowest) and \textit{IQR} (the smallest) of its predictive entropy distribution, and its lowest UCE ($3.118\%$) suggests that its uncertainty prediction is best calibrated.

Despite the good performance of DGP to identify pulsars in this work, a potential constraint that it may suffer from, within the context of image classification, is the dimension of the inputs. In our case for instance, the dimension is $32\times 32$, as opposed to $64 \times 64$ in \cite{guo2019pulsar} and \cite{wang2019pulsar}. As the resolution of an image gets smaller, more information is lost. It would then be interesting to investigate in a future work whether the results presented here can be recovered with an image resolution of $64 \times 64$. It should give us some insights into the impact of the resolution of the inputs on the performance of DGP. However, as shown by \cite{wilson2016stochastic}, Deep Kernel Learning, apart from its scalability, can outperform CNN in classification on dataset like MNIST and CIFAR-10. This is owing to the capacity of the feature extractor (CNN based) it uses, combined with a GP layer. Depending on a given task, the capacity of the model can be adjusted by simply adding or removing convolutional layers.  

\appendix
\section{Gaussian Process (GP)}\label{gaussian_process}
Known as a non-parametric model, the main assumption in a Gaussian Process is that a finite set of real-valued function $\bm{f}$ -- random variables -- evaluated at inputs $\mathbf{X}$ has a joint Gaussian distribution. A Gaussian Process is fully described by the mean $\mu(\mathbf{X})$ and the covariance $K(\mathbf{X}, \mathbf{X'})$ such that \citep{williams2006gaussian} 

\begin{eqnarray}
    \mu(\mathbf{X}) & = & \mathbb{E}[\bm{f}(\mathbf{X})] \nonumber\\
    K(\mathbf{X}, \mathbf{X'}) & = & \mathbb{E}[(\bm{f}(\mathbf{X}) - \mu(\mathbf{X}))(\bm{f}(\mathbf{X'}) - \mu(\mathbf{X'}))].
\end{eqnarray}
A GP prior is then defined as 
\begin{equation}
    \bm{f}(\mathbf{X}) \sim \mathcal{GP}(\mu(\mathbf{X}), K(\mathbf{X}, \mathbf{X'})),
\end{equation}
and setting the mean function to zero for convenience, we have that $p(\bm{f}) = \mathcal{N}(\bm{0}, K(\mathbf{X}, \mathbf{X'}))$.
In reality, the observed functions $\mathbf{y}$ are noisy realizations of the latent functions $\bm{f}$ such that
\begin{equation}
    \mathbf{y} = \bm{f} + \bm{\epsilon},
\end{equation}
where $\bm{\epsilon}$ is a Gaussian noise with a variance $\sigma^{2}$, therefore by adding the variance in quadrature the distribution of the observed functions reads 
\begin{equation}\label{logmarglikely}
    p(\mathbf{y}) = \mathcal{N}(\bm{0}, K(\mathbf{X}, \mathbf{X'}) + \sigma^{2}\mathbf{I}).
\end{equation}
Given that any subset of a collection of random variables which is jointly Gaussian distributed is also a Gaussian distribution, we have the joint distribution of the observations and some test functions $\bm{f}_{*}$ at some test inputs $\mathbf{X}_{*}$ \citep{williams2006gaussian}

\begin{equation}
    \begin{bmatrix}
    \mathbf{y} \\
    \bm{f}_{*}
    \end{bmatrix}  = \mathcal{N}\left(\bm{0}, \begin{bmatrix}
    K(\mathbf{X}, \mathbf{X}) + \sigma^{2}\mathbf{I}) & 
    K(\mathbf{X}, \mathbf{X}_{*}) \\
    K(\mathbf{X}_{*}, \mathbf{X}) & K(\mathbf{X}_{*}, \mathbf{X}_{*})
    \end{bmatrix}\right),
\end{equation}
where $K(\mathbf{X}, \mathbf{X}_{*})$ and $K(\mathbf{X}_{*}, \mathbf{X}_{*})$ are the cross-covariance matrix between training points and test points and the covariance matrix of the test points respectively. Exploiting the property of Gaussian conditionals, we obtain the predictive distribution conditioned on the training data and the test points
\begin{equation}
p(\bm{f}_{*}|\mathbf{y}, \mathbf{X}, \mathbf{X}_{*}) = \mathcal{N}(\mu_{*}, \Tilde{K}),
\end{equation}
where
\begin{equation}\label{invertK}
\mu_{*} = K(\mathbf{X}_{*}, \mathbf{X})[K(\mathbf{X}, \mathbf{X}) + \sigma^{2}\mathbf{I}]^{-1}\>\mathbf{y},
\end{equation}
and
\begin{equation}\label{tildeK}
\Tilde{K} = K(\mathbf{X}_{*}, \mathbf{X}_{*}) - K(\mathbf{X}_{*}, \mathbf{X})
[K(\mathbf{X}, \mathbf{X}) + \sigma^{2}\mathbf{I}]^{-1}K(\mathbf{X}, \mathbf{X}_{*}).
\end{equation}
There is a variety of kernel functions that can be chosen to build the covariance matrix $K(\mathbf{X}, \mathbf{X'})$ but in our analyses we opt for the widely used radial basis function (RBF), given by 
\begin{equation}
    k(x,x') = \sigma_{f}^{2}{\rm exp}\left(-\frac{(x - x')^{2}}{2\ell}\right),
\end{equation}
in which the free parameters $\sigma^{2}_{f}$ and $\ell$ are the signal variance and the length-scale respectively. The latter denotes the scale over which the function $\bm{f}$ significantly changes. These hyperparameters are chosen in such a way as to optimize the log marginal likelihood given in Eq.~\ref{logmarglikely} 
\begin{eqnarray}
    {\rm log}\>p(\mathbf{y}) &=& -\frac{1}{2}\mathbf{y}^{T}[K(\mathbf{X}, \mathbf{X}) + \sigma^{2}\mathbf{I}]^{-1}\mathbf{y}  \nonumber\\
    && - \frac{1}{2}{\rm log}|K(\mathbf{X}, \mathbf{X}) + \sigma^{2}\mathbf{I}| - \frac{n}{2}{\rm log}(2\pi),
\end{eqnarray}
where $n$ is the number of training data points. \cite{williams2006gaussian} provides a practical implementation of a GP regression.
In a classification task, the objective is to predict a category of a given input. Therefore, using GP for classification amounts to \textit{squashing} the latent functions $\bm{f}$ over which the GP prior is defined with a logistic function such as
\begin{equation}
    p(y_{i} = 1 | f_{i}) = \frac{1}{1 + {\rm exp}(-f_{i})},
\end{equation}
where we assume we only have two classes, 0 and 1. In a noise-free GP, the predictive distribution which is conditioned on the training data and the input at some test point is given by
\begin{equation}\label{predictive}
    p(y_{*} = 1 | X_{*}, \mathbf{X}, \mathbf{y}) = \int {\rm d}f_{*}\> p(y_{*} = 1 | f_{*}) p(f_{*} | X_{*}, \mathbf{X}, \mathbf{y}),
\end{equation}
where the latent function evaluated at the test point $f_{*}$ is averaged out. The second term in Eq.~\ref{predictive} is the distribution over the latent functions at a test point and is computed by marginalizing over the latent functions $\bm{f}$ at the training data
\begin{equation}\label{posterior}
    p(f_{*} | X_{*}, \mathbf{X}, \mathbf{y}) = \int {\rm d}\bm{f}\>p(f_{*} | X_{*}, \mathbf{X}, \mathbf{y}, \bm{f})p(\bm{f}|\mathbf{X}, \mathbf{y}),
\end{equation}
where the second term $p(\bm{f}|\mathbf{X}, \mathbf{y})$ is the posterior distribution. Unfortunately, unlike GP regression where the likelihood is Gaussian, the posterior distribution in Eq.~\ref{posterior} is intractable due to the logistic likelihood. The methods, fully described in \cite{williams2006gaussian}, consist of approximating the posterior distribution by a Gaussian distribution, namely Laplace Approximation \citep{williams1998bayesian} and Expectation Propagation \citep{minka2001ep}.
The predictive distribution in a GP involves a matrix inversion of the covariance matrix $K(\mathbf{X}, \mathbf{X'})$. This operation has a cubic complexity $\mathcal{O}(n^{3})$. Therefore, given the nonscalability, dealing with large dataset using GP is quite challenging. To circumvent the issue with the computational complexity, \cite{snelson2005sparse} proposed a sparse GP method by using \textit{"inducing points"} $\mathbf{Z}$ which can be points in the input space corresponding to real valued functions $\mathbf{u}$. The central idea is to select $\mathbf{Z}$ (the number of these inducing variables is $m < n$) , which are fewer than the original data $\mathbf{X}$, such that they capture the characteristics of the functions. Computation of the predictive distribution, by using $\mathbf{Z}$, scales as $\mathcal{O}(m^3)$. \cite{snelson2005sparse}, in their approach, approximated the marginal likelihood by 
\begin{equation}
    p(\mathbf{y}) \sim \mathcal{N}(\mathbf{0}, Q_{nn} + \sigma^{2}\mathbf{I}),
\end{equation}
where $Q_{nn} = {\rm diag}[K_{nn} - K_{nm}K_{mm}^{-1}K_{mn}] + K_{nm}K_{mm}^{-1}K_{mn}$  is low-rank approximation of the original kernel function $K_{nn}$. The overfitting that the setup is prone to, due to the fact that the inducing variables $\mathbf{Z}$ are now part of the hyperparameters to be optimized, led \cite{titsias2009variational} to adopt a different approach which approximates the posterior with a variational distribution which has $\mathbf{Z}$ as part of the variational parameters. \cite{titsias2009variational} then prescribed a lower bound on the marginal likelihood
\begin{equation}
    {\rm log}\>p(\mathbf{y}) \ge \mathcal{N}(\mathbf{0}, Q_{nn} + \sigma^{2}\mathbf{I}) - \frac{1}{2\sigma^{2}}{\rm tr}(K_{nn} - Q_{nn}),
\end{equation}
where the second term involving the trace is known as a regularizer preventing from overfitting.
In \cite{hensman2013gaussian}, they improved on the prescription in \cite{titsias2009variational} by proposing a more scalable bound such that stochastic variational inference is applicable to infer the hyperparameters. 

\acknowledgments

SA acknowledges financial support from the {\it South African Radio Astronomy Observatory} (SARAO). SA is very grateful to Anna Scaife for the recommendations during those fruitful discussions.

\bibliographystyle{JHEP}
\bibliography{pulsar}

\end{document}